# MODERATE GROWTH TIME SERIES FOR DYNAMIC COMBINATORICS MODELISATION

**Luaï JAFF, Gérard H.E. DUCHAMP,**

*LIPN, Paris XIII University*

*99, avenue Jean-Baptiste Clément*

*93430 Villetaneuse, France*

`luai.jaff@gmail.com, gheduchamp@gmail.com`

**Hatem HADJ KACEM**,

*MIRACL, Sfax University*

*Route de l'Aérodrome - BP 559*

*3029 Sfax, Tunisia*

`hatem.hadjkacem@fsegs.rnu.tn`

**Cyrille BERTELLE**,

*LITIS, Le Havre University*

*25 rue Ph. Lebon - BP 540*

*76058 Le Havre Cedex, France*

`cyrille.bertelle@univ-lehavre.fr`

## Abstract

Here, we present a family of time series with a simple growth constraint. This family can be the basis of a model to apply to emerging computation in business and micro-economy where global functions can be expressed from local rules.

We explicit a double statistics on these series which allows to establish a one-to-one correspondence between three other ballot-like strunctures.

**Keywords:** Time series, Complex system, Growth constraint, local rules, Dyck words, Permutations, Codes.

## 1. Introduction

In this paper, we are interested with time series with moderate growth but possibly sudden decay. We will focus ourselves on a very simple model (a "toy model" as physicists may say), the combinatorics of which is completely mastered. This feature is important as one may use simulations and estimates over "all the possible configurations", as it is the case, for example, for other combinatorial models (Cox-Ross-Rubinstein, for instance). The model is that of sequences with integer values and growth bounded by a unit (local rule). Surprisingly, there is one-to-one correspondences between the possible configurations and planar combinatorial objects which are endowed with a special dynamics which we describe here. The structure of the paper is the following. Section 2 presents an applicative economic problem which leads to generate the studied growth time series from local rules. On section 3, we propose a non exhaustive review concerning emerging computation in economic domain and how our work relates corresponding of this body of knowledge. Section 4 develops the dynamic combinatorics computation which leads to establish one-to-one correspondences between three other ballot-like structures. We conclude on section 5.

## 2. From Micro-Economy Local Rules To Dynamic Combinatorics

Our aim is to describe here a **toy-model** of the benefit in the following situation. A capital owner possesses two accounts, say **P** and **R**, **P** is the account where the principal (untouched) capital is deposited. This capital produces a constant return (one unit per unit of time) which is sent to a reserve **R**. From the account **R** can be with drawn arbitrary amounts of money and the account must stay positive.

The possible configurations are described the sequences such that

- $a_1 = 0$
- $a_{i+1} \leq a_i + 1$

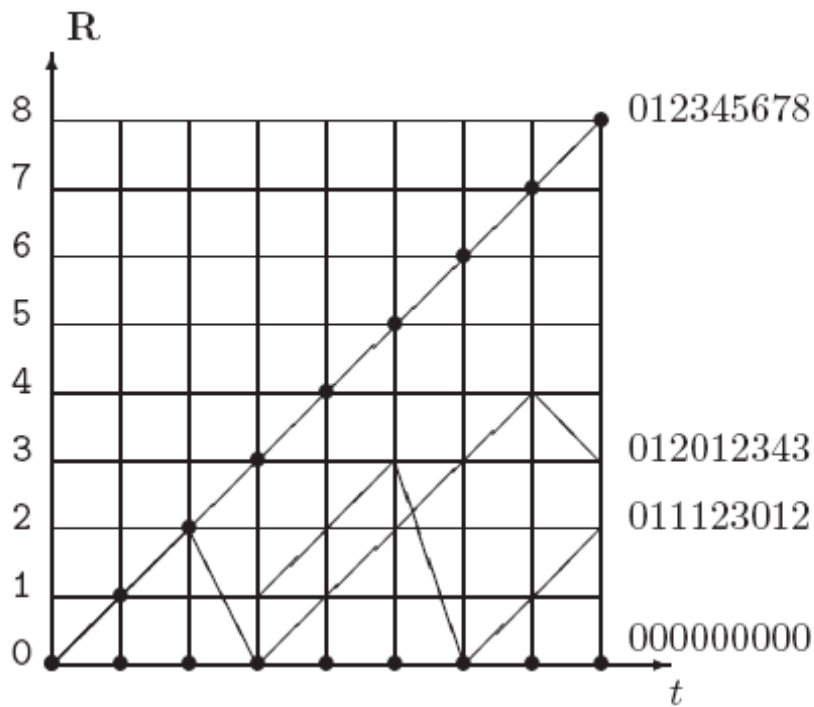

**Figure 1: Maximal, minimal (dotted) and two intermediate trajectories. Their codes are on the right**

In this paper, we build combinatorial structures that allow to modelize and to compute the global behavior of the reserve **R** by some specific functions. We can consider this result as an emergent function from the basic local rules.

**3. Emerging Computations**

Emerging computation is nowadays a thrilling topic which concerns many developments in complex systems modeling. A brief review can allow to classify these emerging computations concerning economic domains in 3 spaces.

The first space is composed of emerging computations which lead to some universal laws. Per Bak's sand pile is concerned by this class [6]. In such model called Self-Organized Criticality, the phenomenon is crossed by transformation which make it evolve by avalanche. The Coton market trade follows such a law. For 1000 small price variations, there are only 100 middle price variations and only 10 major price variations. The general law which characterize such criticality phenomena is an exponential law.

The second space of emerging computation leads to some pattern formations without a complete knowledge of any law. Thomas Schelling's segregation model for urban development is concerned by this class [7]. In such model some local interactionbetween neighbours can lead to self-organized patterns which emerge from the whole interaction systems. Some areas become specialized to some people categories while other areas are devoted to others ones.

The third space of emerging computation described here, leads to some global functions expressions. It is typically what we will describe in our problem. The local rules concerned by the proposed economic toy-model will lead to define combinatorics structures allowing to compute a functional global approach. The detailed computation is describe in the following

## 4. Dynamics Combinatorics Computation

### 4.1 Trajectories and Codes

We can define the trajectories of our model by sequences (codes) $a_1 a_2 a_3 ... a_n$ such that

- $a_1 = 1$
- $a_{j+1} \leq a_j + 1$

**Example :** For n = 4, we have 14 codes as described in the following table.

| numbers | codes |
|---|---|
| 1 | 1111 |
| 2 | 1112 |
| 3 | 1121 |
| 4 | 1122 |
| 5 | 1123 |
| 6 | 1211 |
| 7 | 1212 |
| 8 | 1221 |
| 9 | 1222 |
| 10 | 1223 |
| 11 | 1231 |
| 12 | 1232 |
| 13 | 1233 |
| 14 | 1234 |

We remark that we have 5 codes which end by 1 or 2, 3 codes ending by 3 and one code ending by 4. Now if one sets $l(n,k)$ to be the number of codes ending by $k-1$, one can check that

- $l(n,0) = l(0,n) = 0 \ (\forall n \geq 1)$
- $l(0,0) = 1$ (the void sequence)
- $l(n,k) = \sum_{j \geq k+1} l(n-1, j)$

whence the easy computed table of the first values

| N\k | 0 | 1 | 2 | 3 | 4 | 5 | 6 | 7 |
|---|---|---|---|---|---|---|---|---|
| 0 | 1 | 0 | 0 | 0 | 0 | 0 | 0 | 0 |
| 1 | 0 | 1 | 0 | 0 | 0 | 0 | 0 | 0 |
| 2 | 0 | 1 | 1 | 0 | 0 | 0 | 0 | 0 |
| 3 | 0 | 2 | 2 | 1 | 0 | 0 | 0 | 0 |
| 4 | 0 | 5 | 5 | 3 | 1 | 0 | 0 | 0 |
| 5 | 0 | 14 | 14 | 9 | 4 | 1 | 0 | 0 |
| 6 | 0 | 42 | 42 | 28 | 14 | 5 | 1 | 0 |
| 7 | 0 | 132 | 132 | 90 | 48 | 20 | 6 | 1 |

The values for $n, k \geq 1$ can be even more easily computed with the (subdiagonal) local rule described by **West** + **North** = **result**. For instance, we remark that 9 + 5 = 14.

| 1 | 0 | 0 | 0 | 0 | 0 | 0 |
|---|---|---|---|---|---|---|
| 1 | 1 | 0 | 0 | 0 | 0 | 0 |
| 1 | 2 | 2 | 0 | 0 | 0 | 0 |
| 1 | 3 | 5 | 5 | 0 | 0 | 0 |
| 1 | 4 | 9 | 14 | 14 | 0 | 0 |
| 1 | 5 | 14 | 28 | 42 | 42 | 0 |

Remark that the preceding table gives the mirror images of the lines of the previous double statistics.

## 4.2 Permutations

We say that a permutation $\pi$ of n letters has an increasing subsequences of length $k$ if there are positions

$$1 \leq i_1 < i_2 < i_3 < ... < i_k \leq n$$

such that

$$\pi(i_1) < \pi(i_2) < \pi(i_3) < ... < \pi(i_k)$$

For example

$$\pi = \begin{pmatrix} 1 & 2 & 3 & 4 & 5 \\ 5 & 3 & 4 & 1 & 2 \end{pmatrix}$$

has increasing subsequences of length 2, at points {2,3} as well as at positions {4,5}. Let $\pi_2(n)$ be the number of permutations of n letters that have no increasing subsequences of length $>2$. By direct enumeration we obtain the following table.

| n | 0 | 1 | 2 | 3 | 4 | 5 | 6 | 7 | 8 |
|---|---|---|---|---|---|---|---|---|---|
| $\pi_2(n)$ | 1 | 1 | 2 | 5 | 14 | 42 | 132 | 429 | 1430 |

**Proposition 1.** $|\pi_2(n)| = C_n$, where $\pi_2(n)$ the number of permutations of n letters that have no increasing subsequences of length $>2$ and $C_n$ is the n-th Catalan number.

$$C_n = \frac{1}{n+1}\binom{2n}{n}$$

## 4.3 Young Tableaux

**Definition 1.** A **partition** of n, written $\lambda \triangleright n$, is a sequence,

$$\lambda = (\lambda_1, \lambda_2, \lambda_3, ..., \lambda_k)$$

such that the $\lambda_i$ are decreasing (weakly) and $\sum_{i=1}^{k} \lambda_i = n$.

Let $\lambda = (\lambda_1, \lambda_2, ..., \lambda_n) \triangleright n$. Then the **Ferrers diagram**, or **shape**, of $\lambda$ is an array of n-squares into k left-justified rows with row $i$ containing $\lambda_i$ squares for $1 \leq i \leq k$. For example, the partition (4,3,1)

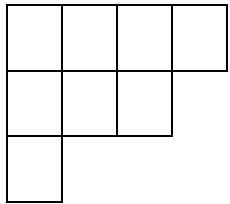

Let $\lambda$ be as above. A **Young tableau** of shape $\lambda$, is an array obtained by replacing the squares of the shape of $\lambda$ by a bijection with the numbers 1,2,...,n. A tableau $T$ is said to be a **standard Young tableau** if the rows and columns are increasing sequences. For example below the tableau is standard

| 1 | 2 | 3 | 5 |
|---|---|---|---|
| 4 | 6 | 7 |   |
| 8 |   |   |   |

A **standard Young tableau of two lines** is of the shape $\lambda(l_1, l_2) \triangleright n$ where $l_1 \geq l_2 > 0$. Let $f^{(l_1, l_2)}$ be the number of standard tableaux of two lines. We have

$$f^{(l_1, l_2)} = \frac{l_1 - l_2 + 1}{l_1 + 1} \binom{l_1 + l_2}{l_1}$$

**Proposition 2.** If $l_1 = l_2$ then

$$f^{(l_1, l_1)} = \frac{1}{l_1 + 1} \binom{2l_1}{l_1}$$

which is th n-th Catalan number $C_n$.

In general, we can represent a Young tableaux of two (equal) lines as follows :

| . | . | . | . | 2n |
|---|---|---|---|----|
| 1 | . | . | . | k  |

Then

$$2n - k = h$$

will be the second parameter of the tableau.

**Example :** A Young tableau of size six and height two

| 3 | 5 | 6 |
|---|---|---|
| 1 | 2 | 4 |

Then

$$6 - 4 = 2$$

we remark that 2 is the last letter of

$$\sigma_2(3) = 312$$

## 4.4 Dyck Words and Dyck Paths

Let **w** be a word and **a** a letter, the length of **w** will be denoted by $|w|$, and the number of occurrences of **a** in **w** by $|w|_a$. We denote the empty word by $\varepsilon$. If **w** = **uv**, then **u** is a prefix of **w**.

**Definition 2.** A Dyck word **w** is a word over the alphabet $\Sigma = \{0,1\}$ with the following properties :

- For each prefix of $u$ of $w$, $|u|_1 \geq |u|_0$
- $|w|_1 = |w|_0$

A Dyck path is a path in the first quadrant, which begins at the origin $(0,0)$, ends at $(2n,0)$. A Dyck path consists **North-east** and **South-east** steps. The number of Dyck paths of length 2n is the Catalan number

$$C_n = \frac{1}{n+1}\binom{2n}{n}$$

and thus,

$$\sum_{n \geq 0} C_n x^n = \frac{1-\sqrt{1-4x}}{2x}$$

Now, we can refine the number of Dyck words with respect to a parameter like k which is the number of factors in Dyck words. The number of Dyck words of length 2n which decompose into $k$ (irreducible Dyck) factors is exactly $l(n,k)$ and their sum (over $k$) equal to the Catalan numbers. For example

aaaabbbb

is a Dyck word of length 4 which has one factor.

## *4.5 Bijections*

In this chapter we will describe the links between some combinatorial famillies and we try to give certain properties that help us to understand the connection.

$$Codes \leftrightarrow \sigma_2(n) \leftrightarrow Youngtableaux \leftrightarrow Dyckwords$$

**Theorem 1.** Let $\Phi = \{a\} \cup \Phi^+$ be a data structure with a bi-variate statistics

$$l : \Phi \to N^2$$
$$s \to l(s) = (n, k)$$

such that

$$l(\Phi^+) \subset N^+ \times N$$

Suppose that

We suppose that there exist a function $d : \Phi^+ \to \Phi$ such that

1. $d : \Phi_n \to \Phi_{n-1}$ ($\Phi_n = (pr_1 \circ l)^{-1}(n)$)

2. $\phi : \Phi_n \to \Phi_{n-1} \times N^+$

    $s \to (d(s), k)$ is injective

3. define $\pi = pr_2 \circ l$. For all $s \in \Phi$ we define his code by

$$\chi(s) = (\pi(d^{n-1}(s)), \pi(d^{n-2}(s)), \cdots \pi(d(s)), \pi(s))$$

then $\chi$ is injective.

For example to pass a code 1122 to $\sigma_2(4)$, $f^{(4,4)}$ and Dyck word of length 8 which decompose into one factor.

$$\sigma_2(2) : 1 \to 12 \to 312 \to 3412$$

Dyck word : ab → aabb → abaabb → aabaabbb

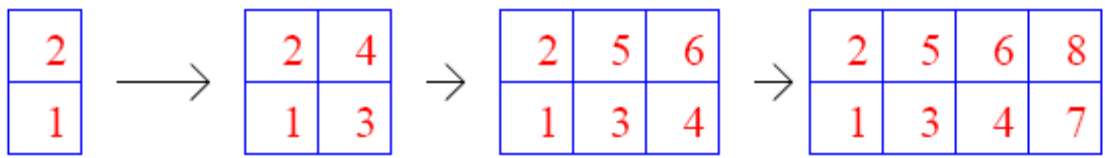

**Figure 2: Young tableau**

## 5. Conclusion

We have presented an toy-model economic behaviour based on local rules and we propose some global function expression which can be also described by three combinatorics structures. By this application, we point out a one-to-one correspondence between three other ballot-like structures. The innovative aspect of this paper deals with a constructive development of the involved bijections.

## References

[1]   B.E. Sagan, *The symmetric group*, 1991 .

[2]   C. Schensted, *Longest increasing and decreasing subsequences*, Canad.J.Math, 13(1961), 179-191.

[3]   M.P. Schützenberger, *Quelques remarques sur une construction de Schensted*, Math. Scand., 12(1963), 117-128.

[4]   Herbert S. Wilf, *Ascending subsequences of permutations and the shape of tableaux*, Journal of Combin. theory, series A 60(1992), 155-157.

[5]   Herbert S. Wilf, *The computer-aided discovery of a theorem about Young tableaux*, J. Symbolic Computation, series 20 (1995), 731-735.

[6]  Per Bak, *How nature works - the science of self-organized criticality*, Springer Verlag, 1996.

[7]   Thomas Schelling, *Dynamic models of segregation*, J. of Math. Sociology, vol. 1, 1971.